\documentclass[reprint,prl, superscriptaddress]{revtex4-1}
\usepackage{amssymb}
\usepackage{graphicx}
\usepackage{amsmath}
\usepackage{natbib}
\usepackage{siunitx}
\usepackage{textcomp}
\usepackage{bm}
\usepackage{color}
\usepackage[hidelinks,colorlinks=true,urlcolor=blue,linkcolor=blue,citecolor=blue]{hyperref}

\def\ba{{\boldsymbol a}}
\def\brho{{\boldsymbol \brho}}
\def\bk{{\boldsymbol k}}

\def\br{\boldsymbol{r}}

\def\bJ{{\boldsymbol J}}
\def\bK{{\boldsymbol K}}

\def\la{\langle}

\def\calH{\mathcal{H}}

\def\calA{\mathcal{A}}

\def\pa{\partial}
\def\nn{\nonumber}

\def\la{\langle}
\def\ra{\rangle}
\begin{document}

\title{Intrinsic Anomalous Hall Effect in a Bosonic Chiral Superfluid}
\author{Guan-Hua Huang}
\affiliation{Department of Physics, Southern University of Science and Technology, Shenzhen, 518055, China}
\affiliation{Shenzhen Institute for Quantum Science and Engineering, Southern University of Science and Technology, Shenzhen 518055, China.}
\author{Zhi-Fang Xu}
\email{xuzf@sustech.edu.cn}
\affiliation{Department of Physics, Southern University of Science and Technology, Shenzhen, 518055, China}
\affiliation{Shenzhen Institute for Quantum Science and Engineering, Southern University of Science and Technology, Shenzhen 518055, China.}
\affiliation{International Quantum Academy, Shenzhen 518048, China.}
\affiliation{Guangdong Provincial Key Laboratory of Quantum Science and Engineering, Southern University of Science and Technology, Shenzhen 518055, China.}
\author{Zhigang Wu}
\email{wuzg@sustech.edu.cn}
\affiliation{Shenzhen Institute for Quantum Science and Engineering, Southern University of Science and Technology, Shenzhen 518055, China.}
\affiliation{International Quantum Academy, Shenzhen 518048, China.}
\affiliation{Guangdong Provincial Key Laboratory of Quantum Science and Engineering, Southern University of Science and Technology, Shenzhen 518055, China.}

\date{\today }
\begin{abstract}
The anomalous Hall effect has had a profound influence on the understanding of many electronic topological materials but is much less studied in their bosonic counterparts. 
We predict that an intrinsic anomalous Hall effect exists in a recently realized bosonic chiral superfluid,  a $p$-orbital Bose-Einstein condensate in a 2D hexagonal boron nitride optical lattice [X. Wang {\it et al.}, \href{https://www.nature.com/articles/s41586-021-03702-0} {\textcolor{blue}{Nature (London) {\bf 596}, 227 (2021)}}]. We evaluate the frequency-dependent Hall conductivity within a multi-orbital Bose-Hubbard model that accurately captures the real experimental system. We find that in the high frequency limit, the Hall conductivity is determined by finite loop current correlations on the $s$-orbital residing sublattice, the latter a defining feature of the system's chirality. In the opposite limit, the dc Hall conductivity can trace its origin back to the non-interacting band Berry curvature at the condensation momentum, although the contribution from atomic interactions can be significant. We discuss available experimental probes to observe this intrinsic anomalous Hall effect at both zero and finite frequencies. 
\end{abstract}
\maketitle

{\it Introduction.}---The capacity of ultracold atomic gases as quantum simulators of more complex condensed matter systems owes to the plethora of experimental tools available to provide these neutral atoms with solid-state-like settings~\cite{2012Bloch,2014Georgescu,zhai_2021}. Well-known examples include optical lattice potentials~\cite{1998Jaksch,2008Bloch}, synthetic spin-orbit couplings~\cite{2011Lin,2013Galitski,Zhai_2015} and artificial gauge fields~\cite{2011Dalibard,2012Leblanc}. Motivated by the desire to simulate electronic systems for which orbital degrees of freedom are essential~\cite{2000Tokura}, experimentalists have begun to load atoms onto higher Bloch bands of optical lattices of various crystal structures~\cite{2016Kock,2016XiaopengLi,2007Muller,2010Wirth,2012Soltan,2013Olschlager,2015Kock,2016Sengstock,2021Hachmann,2021Vargas,2021Jin,2021Wang}. With the choice of bosonic atoms, this experimental approach has also led to the creation of novel atomic superfluids~\cite{2010Wirth,2012Soltan,2013Olschlager,2015Kock,2021Jin,2021Wang} which, among other things, exhibit a spontaneous breaking of time-reversal symmetry (TRS) due to atomic interactions.

The recently realized $p$-orbital Bose-Einstein condensate in a 2D boron nitride optical lattice~\cite{2021Wang} is a particularly interesting example of such superfluids. A prominent feature of this system is that it acquires a macroscopic angular momentum in the absence of any artificial magnetic field or rotation of the trapping potential,  and displays a chirality akin to that of superfluid $^3$He~\cite{2012Walmsley,2013Ikegami}. Furthermore, it contains a topologically non-trivial quasi-particle band structure as well as gapless edge states, both reminiscent of similar concepts  in fermionic topological materials~\cite{2013Bernevig}. Because the Hall response often plays a fundamental role in understanding and characterizing these topological materials~\cite{1982TKNN,1988Haldane,2016Liu}, it is only natural to ask if an anomalous Hall effect (AHE), quantum or otherwise, exists in their newly discovered bosonic counterpart. 

In electronic materials, the intrinsic (scattering-free)  AHE and its quantum version are both well understood in terms of the geometrical properties of the topologically non-trivial band structures~\cite{2010Nagaosa,2010Xiao}. In this framework, the dc Hall conductivity is determined by the summation of Berry curvatures of the occupied Bloch states, weighted by the Fermi-Dirac distribution function~\cite{2010Xiao}. This means that  AHE may also occur in bosonic quantum gases if the atoms are excited to states having non-zero Berry curvature, either in a non-equilibrium situation~\cite{2004Dudarev,2015Li} or at finite temperatures~\cite{2011Bijl,2018patucha}. At zero temperature, however,  earlier studies found vanishing dc Hall responses in bosonic models for both topologically trivial~\cite{2015Li} and non-trivial~\cite{2018patucha} band structures.

In addition to the dc Hall response, the ac anomalous Hall effect, i.e.,  the frequency-dependent Hall response, has been widely studied in chiral superconductors in the context of the Kerr effect~\cite{2012Taylor,2009Lutchyn,2016Kallin,2019Brydon,Denys2021}. There, the Hall response originates not from a topologically non-trivial band structure but rather from the TRS-breaking superconductivity. In particular, a remarkable relation is found between the high-frequency Hall response and the loop current correlations~\cite{2019Brydon}. The dc Hall conductivity in those systems, however, cannot be directly related to the Berry curvatures of the electronic bands~\cite{2009Lutchyn}. 

In this Letter, we show that an intrinsic AHE indeed exists in the bosonic chiral superfluid at zero temperature, for both zero and finite frequencies. Fascinatingly, both of the physical mechanisms aforementioned are at play in our system. At high frequencies, we found a striking similarity to the chiral superconductors in that the Hall response appears as a consequence of the chirality. The dc Hall conductivity, on the other hand, can still be understood in terms of the Berry curvature of the non-interacting bands, although the effect of atomic interactions needs to be accounted for. We will discuss experimental methods uniquely suitable to detecting the AHE in this ultracold atomic system.

\begin{figure}[tbp]
\includegraphics[width=8.7cm]{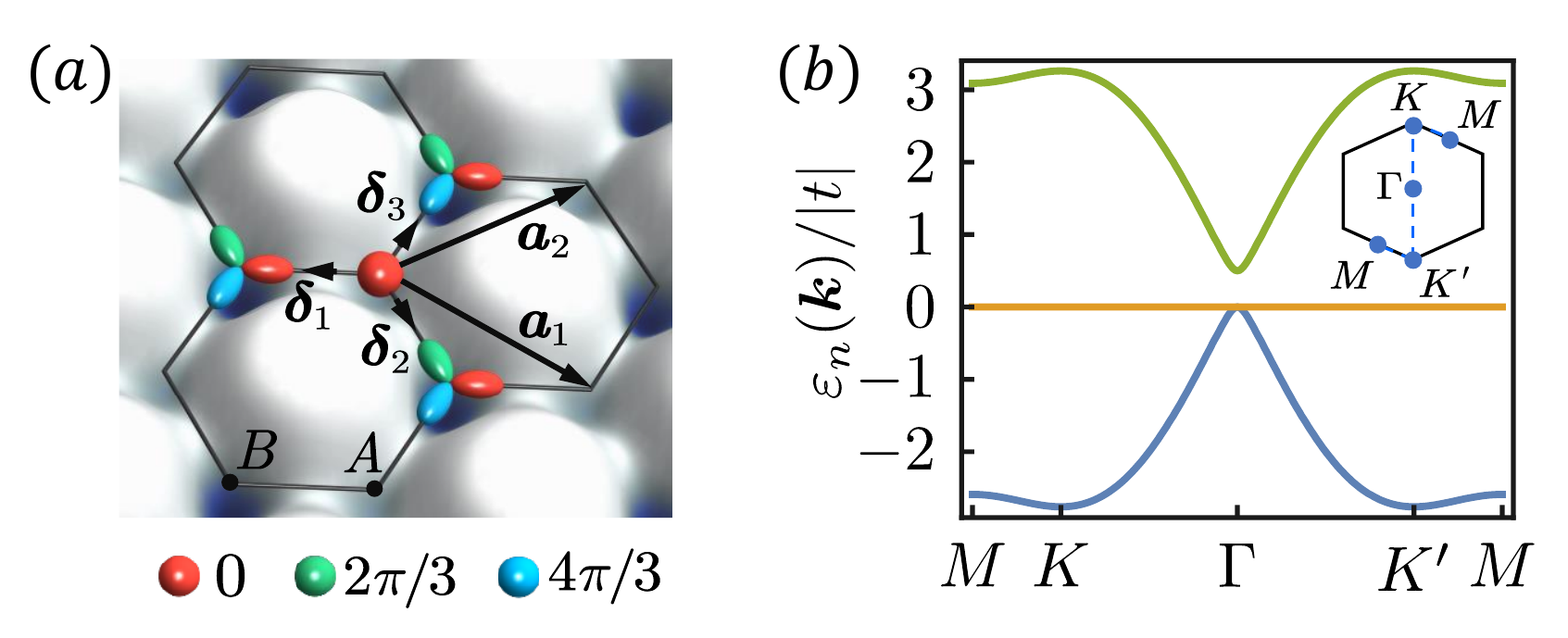}
\caption{\label{fig1} (a) Illustration of the boron nitride lattice potential, where the atoms reside in the $s$-orbital on $A$ sites and in the two degenerate $p$-orbitals on $B$ sites. The colors indicate the phase winding of the $s$ and $p$-orbital wave functions. (b) Non-interacting band dispersions of the BN lattice described by $ h_0$ in Eq.~(\ref{h0}), for  $\epsilon_s-\epsilon_p = 0.5 |t|$. Note that there is a band touching point at $\Gamma$. The inset depicts the first Brillouin zone and shows the path along which the dispersions are plotted.}
\end{figure}

{\it Bosonic chiral superfluid}--- The atomic chiral superfluid is realized in  a quasi-2D $^{87}$Rb gas confined in a hexagonal boron nitride (BN) optical lattice~\cite{2021Wang}. The BN lattice is formed by superimposing two sets of triangular lattice potential $A$ and $B$, spanned by primitive vectors $\ba_1 = a(\sqrt{3}/2,-1/2)$ and $\ba_2 = a(\sqrt{3}/2,1/2)$, where $a$ is the lattice constant (see Fig.~\ref{fig1}(a)). The local potential wells on $B$ sublattice sites are deeper than those on $A$, such that the energy of the $s$-orbital on $A$ sites is comparable to that of the two degenerate  $p$-orbitals on $B$. In such a case, these three orbitals form a subspace in which an effective multi-orbital Bose-Hubbard model can be constructed to describe the system with the Hamiltonian $\hat H = \hat H_0 + \hat H_{\rm int} $, where the kinetic and interaction energy are given by~\cite{SM} 
\begin{displaymath}
 \hat H_{0}=\sum_{\br}\epsilon_{s}\hat s_{\br}^{\dagger}\hat s_{\br}+\sum_{\br,\alpha}\epsilon_{p}\hat p_{\alpha,\br}^{\dagger}\hat p_{\alpha,\br} +\sum_{\br,\alpha, l}(t_{l}^\alpha \hat s_{\br}^{\dagger}\hat p_{\alpha,\br+\boldsymbol{\delta}_{l}}+h.c.)\label{eq:tightbinding1} \nn
\end{displaymath}
and
\begin{displaymath}
\hat H_{\rm int}=\frac{U_{s}}{2}\sum_{\br}\hat s_{\br}^{\dagger}\hat s_{\br}^{\dagger}\hat s_{\br}\hat s_{\br}+\frac{U_{p}}{2}\sum_{\br,\alpha\alpha'\beta\beta'}\hat p_{\alpha',\br}^{\dagger}\hat p_{\beta',\br}^{\dagger}\hat p_{\beta,\br}\hat p_{\alpha,\br} \nn
\end{displaymath}
respectively. Here in the second term of $\hat H_{\rm int}$ only the angular momentum conserving processes are retained in the summation. In addition, throughout the paper, summation over the lattice vector $\br$ is understood to be restricted to $A$ sublattice for $s$-orbital and to $B$ sublattice for $p$-orbitals. Thus, $\hat s_{\br}^{\dagger}$ creates a $s$-orbital atom with energy $\epsilon_s$  on lattice site $\br \in A$. The two degenerate $p$-orbitals with energy $\epsilon_p$, created by $\hat p_{\alpha,\br}^{\dagger}$ ($\alpha=\pm$) on site $\br \in B$, are eigenstates of the rotation operator and  thus time-reversed counterparts of each other. The nearest neighbor hopping parameter along the vector $\bm\delta_l$ ($l=1,2,3$) can be written as $t_l^+ =t_l^{-*}= t e^{i(l-1)2\pi/3}$ due to the phase winding of the $p$-orbitals and the $C_3$ symmetry of the lattice. The on-site interaction strengths for $s$ and $p$-orbitals, $U_s$ and $U_p$ respectively, are both proportional to the scattering length of the atoms. 

The non-interacting band dispersion $\varepsilon_n(\bk) $ ($n=0,1,2$), as shown in Fig~\ref{fig1}(b), can be solved analytically by diagonalizing 
\begin{align}
h_0(\bk) \equiv \begin{pmatrix}\epsilon_{s}  & f^+_\bk & f^-_\bk\\
& \epsilon_{p} & 0\\
h.c. &  & \epsilon_{p}
\end{pmatrix},
\label{h0}
\end{align} 
where $f_{\bm{k}}^\pm=\sum_{l}t^{\pm}_{l}e^{i\bm{k}\cdot\bm{\delta}_{l}}$. Since the Hamiltonian preserves the time-reversal symmetry, the dispersion has the property $\varepsilon_n(\bk) = \varepsilon_n(-\bk)$.   In addition, the local minima of the bottom band occur at the time-reversed pair of momenta $\bK$ and $\bK'$, as can be seen from the band dispersion $\varepsilon_0(\bk) = \frac{1}{2}[\epsilon_{s}+\epsilon_{p}-\sqrt{(\epsilon_{s}-\epsilon_{p})^{2}+4(|f^+_{\boldsymbol{k}}|^{2}+|f^-_{\boldsymbol{k}}|^{2})}]$. 

At zero temperature, a mean-field analysis shows that the presence of weak interactions leads to the condensation of atoms into either $\bK$ or $\bK'$ in the momentum space, spontaneously breaking the time-reversal symmetry. It can be further shown by symmetry considerations that the $\bK$ ($\bK'$) ground state corresponds to the condensation of atoms into pure $p_+$ ($p_-$) orbital on $B$ sublattice sites~\cite{SM}. Taken together, the condensate wave function associated with $\bK=\frac{4\pi}{3a}(0,1)$ condensation can be written as 
\begin{align}
 \la \hat \psi_\bk \ra=  \sqrt{N}(\cos{\xi}, \sin{\xi},0  )^T \delta_{\bk,\bK}.
 \label{cwf}
 \end{align}  
Here we adopt a spinor notation $\hat \psi_\bk \equiv (\hat s_{\bk},\hat p_{+,\bk} , \hat p_{-,\bk} )^T $ where $\hat s_{\bk} = \frac{1}{\sqrt{N_u}}\sum_{\br} e^{-i\bk \cdot \br} \hat s_{\br}$ with $N_u$ as the number of unit cells; $N$ is the number of atoms,  and $\xi$ is determined by minimizing the Gross-Pitaevskii energy~\cite{SM}. The key property that distinguishes this multi-orbital superfluid from many of the previously realized atomic superfluids is that it is globally chiral,  carrying a total angular momentum of $\la {\hat L}_z\ra = \hbar \sum_{\br} \la \hat p_{+,\br}^{\dagger}\hat p_{+,\br} -\hat p_{-,\br}^{\dagger}\hat p_{-,\br}   \ra = N\hbar (\sin{\xi})^2 $. 

\begin{figure}[tbp]
\includegraphics[width=8.7cm]{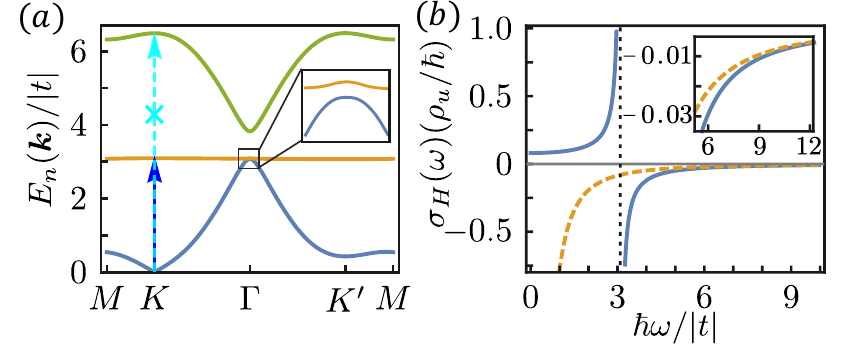}
\caption{\label{fig2} (a) Bogoliubov spectrum of the chiral superfluid. The inset shows an expanded view of a small gap opening at the $\Gamma$ point, which renders the Bogoliubov excitations topological~\cite{2020Zhou,2021Wang}. The arrow with the solid line indicates the transition from the ground state to the first excitation band at $\bK$ in the presence of an ``electric field"; the arrow with the dashed line indicates the absence of such transitions to the second excitation band. (b) Frequency-dependent Hall conductivity. The dashed line shows the asymptotic behavior of $\sigma_H(\omega)$ given by Eq.~(\ref{asym}). The inset is an expanded view of this asymptotic behavior . Here $\epsilon_s - \epsilon_p = 0.5|t|$, $\rho_u U_s = 1.2|t|$ and $\rho_u U_p = 0.72|t|$.}
\end{figure}

{\it Anomalous Hall effect}--- We first demonstrate the AHE by evaluating the frequency-dependent Hall conductivity using the Kubo formula 
\begin{align}
\sigma_{H}(\omega)\equiv  \frac{1}{\mathcal A\omega}{\rm Im}\chi^J_{x,y}(\omega),
\label{Kubo}
\end{align}
where $\mathcal A$ is the  system's area, and $\chi^J_{x,y}(\omega)$ is the Fourier transform of the retarded current-current correlation function 
$\chi^J_{x,y}(t-t') = -i\hbar^{-1} \theta(t-t') \la {[\hat J_x(t), \hat  J_y(t')]} \ra$. Here the total current operator takes the familiar form of
\begin{align}
\hat \bJ&=  \frac{i}{\hbar}\sum_{\bm{r},l} \left [ \boldsymbol{\delta}_{l}(t_{l}^+\hat s_{\bm{r}}^{\dagger}\hat p_{+,\bm{r}+\bm{\delta}_{l}}+t_{l}^{-}\hat s_{\bm{r}}^{\dagger}\hat p_{-,\bm{r}+\bm{\delta}_{l}})-h.c.\right ]. 
\label{current}
\end{align}
To calculate the Hall conductivity we need to determine the collective excitations, which can be found by solving the Bogliubov-de Gennes equation 
\begin{align}
\tau_z\calH_B(\bk) V_{n}(\bk) = E_{n}(\bk) V_{n}(\bk).
\label{Bogo}
\end{align}
Here $\tau_z = \sigma_z\otimes I$, where $\sigma_z$ is the Pauli matrix and $I$ is the $3\times 3$ identity matrix; $E_n(\bk)$ is the Bogoliubov spectrum, $V_n(\bk)$ is the corresponding amplitude with the normalization $V_m^\dag(\bk)\tau_z V_n(\bk) = \tau_{z,mn}$ and the matrix $\calH_B(\bk)$ (assuming the $\bK$ condensation) is given by
\begin{align}
\mathcal{H}_B(\bk) = \left(\begin{matrix}
h_0(\bk) + g - \mu I  & g' \\
g'& h_0^*(2\bK -\bk) + g-\mu I 
\end{matrix}   \right)
\end{align}
 where $g ={\rm diag} ( 2U_s \rho_u \cos^2\xi , 2U_p \rho_u \sin^2\xi  ,2U_p \rho_u \sin^2\xi  )$ and $g'= {\rm diag} ( U_s \rho_u \cos^2\xi , U_p \rho_u \sin^2\xi  ,0  )$ are diagonal matrices accounting for the atomic interactions, $\rho_u$ is the number of atoms per unit cell and  $\mu$ is the chemical potential. We note that $\xi$ and $\mu$ in $\mathcal{H}_B(\bk) $ are calculated previously when obtaining the condensate wave function in Eq.~(\ref{cwf}). Shown in Fig.~\ref{fig2}(a) is a typical plot of the Bogoliubov spectrum obtained from Eq.~(\ref{Bogo}). As usual the full solutions to Eq.~(\ref{Bogo}) include both the positive ($n= 0,1,2$)  and negative ($n= \bar 0,\bar 1,\bar 2$)  branches, but only the former represent the physical excitations and are shown in the figure. Importantly, the gapless solution of Eq.~(\ref{Bogo}) reproduces the condensate wave function, namely $V_0(\bK) = (1/\sqrt{2}) (\Phi, -\Phi^*)^T$ where $\Phi = ( \cos{\xi},\sin{\xi},0)$.
 
Within the Bogoliubov theory, the Hall conductivity defined in Eq.~(\ref{Kubo}) is given by~\cite{SM}
\begin{align}
 \sigma_H(\omega) =&\frac{N}{\calA\omega\hbar^2}{\rm Im}\sum_{n=1,2}\left[\frac{J_{x,0n}(\bK)J_{y,n0}(\bK)}{\hbar\omega - E_n(\bK)+E_0(\bK)+i0^+} \right.\nn \\
 &\left. -\frac{J_{y,0n}(\bK)J_{x,n0}(\bK)}{\hbar\omega +E_n(\bK)- E_0(\bK)+i0^+}\right ],
 \label{Hall}
\end{align} 
where $J_{i,0n}(\bK) = \sqrt{2}V^\dag_0(\bK)\pa_i\calH_B(\bK)V_n(\bK)$ with $\pa_i \equiv \pa /\pa k_i$. The Hall conductivity calculated from Eq.~(\ref{Hall}) is plotted in Fig.~\ref{fig2}(b). From the expression we see that the finite Hall conductivity comes from the inter-band transitions at momentum $\bK$ when an ``electric field" is applied to the system. Here it is critical that the condensation occurs at a momentum for which the velocity operator $\nabla_\bk \calH_B(\bk)$ is finite; otherwise no inter-band transitions would take place and hence no anomalous Hall effect~\cite{2015Li,2018patucha}.  Another notable aspect is the absence of  transitions to the second excitation band as a result of the selection rule $J_{i,02}(\bK) = 0$. This explains why only one singularity, located at $\hbar\omega = E_1(\bK)- E_0(\bK)$, appears in the plot of $\sigma_H(\omega)$  in Fig.~\ref{fig2}(b). This selection rule reflects some of the general properties of the high-symmetry point $\bK$ in the Brillouin zone, valid beyond our tight-binding model~\cite{SM}.  
To understand other features of $\sigma_H(\omega)$ and the physical reasons underlying the AHE, we next analyze both the high and low frequency limit of the Hall response. 

\begin{figure}[btp]
\includegraphics[width=8.7cm]{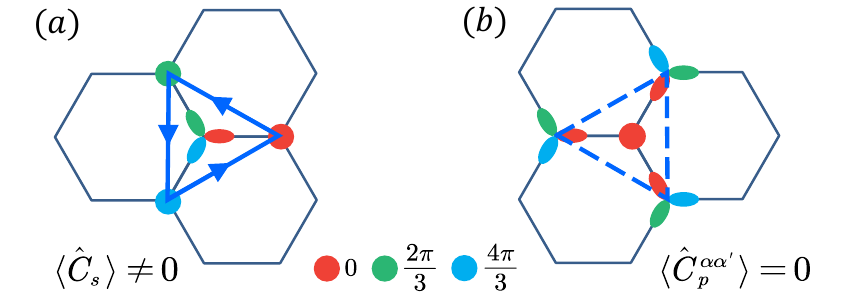}
\caption{\label{fig3} (a) Finite loop current correlation on the $s$-orbital residing $A$ sublattice. (b) Vanishing loop current correlations on the $p$-orbital residing $B$ sublattice. }
\end{figure}

{\it High frequency Hall response and loop current correlations.}---In the high frequency limit $\hbar \omega \gg  |t|$, it can be shown from the Kubo formula  that~\cite{1993Shastry}
\begin{align}
\sigma_H(\omega) = \frac{1}{\hbar\calA \omega^2}  {\rm Im}\la [\hat J_x, \hat J_y] \ra + O\left(\frac{1}{\omega^4}\right ).
\end{align}  
Using Eq.~(\ref{current}) to evaluate the above commutator we find in this limit
\begin{align}
 \sigma_H(\omega) = \frac{ |t|^2}{ 3N_u\hbar^3\omega^2} \bigg( \la \hat C_s \ra + \sum_{\alpha\alpha'} \la\hat C_p^{\alpha\alpha'}\ra \bigg), 
 \label{sigmaloop}
\end{align}  
 where $\hat {C}_{s}$ and $\hat {C}^{\alpha\alpha'}_{p}$ are the loop current correlators 
\begin{align}
\hat {C}_{s}=i & \sum_{\bm{r}, \la l l' \ra} \big (e^{i\pi}\hat s_{\br-{\bm \delta}_{l'}}^{\dagger}\hat s_{\br-{\bm \delta}_l} -h.c. \big );   \\
\hat {C}^{\alpha\alpha'}_{p}=i & \sum_{\bm{r}, \la l l' \ra} \big (e^{i\theta_{ll'}^{\alpha\alpha'}}\hat p_{\alpha',\br+{\bm \delta}_{l'}}^{\dagger}\hat p_{\alpha,\br+{\bm \delta}_l} -h.c. \big ).
\end{align}
Here $\la ll'\ra$ denotes the summation over the cyclic permutations $( 1\,2) $;$(2\,3) $;$(3\,1)$, $\theta_{ll'}^{++} = -\theta_{ll'}^{--} = -\bK\cdot({\bm\delta}_l - {\bm \delta}_{l'})$ and
$\theta_{ll'}^{+-} = -\theta_{ll'}^{-+} = -\bK\cdot({\bm\delta}_l + {\bm \delta}_{l'})$. We note that $\hat C_s$ is defined on sets of three nearest-neighbor  $A$ sites, each set encircling a $p$-orbital residing $B$ site as illustrated in Fig.~\ref{fig3}(a); similarly $\hat C^{\alpha \alpha'}_p$ are defined along the triangles encircling the $s$-orbital residing $A$ sites.
  If we include nearest neighbor hoppings within each sublattice, these correlators determine the so-called loop current via
$
\hat J_{ loop} = -a \big(|t_s| \hat C_s  +  \sum_{\alpha\alpha'} |t_p^{\alpha\alpha'}| \hat C_p^{\alpha\alpha'}\big)
$,
where $|t_s|$ and $|t_{p}^{\alpha\alpha'}|$ are the nearest neighbor hopping strengths within $A$ and $B$ sublattices respectively~\cite{SM}. Relations analogous to Eq.~(\ref{sigmaloop}) were first discussed in models of spin-singlet chiral d-wave superconductors~\cite{2019Brydon}.

Now Eq.~(\ref{sigmaloop}) allows us to see the connection between the chirality of the system and the high frequency Hall response. As discussed earlier, the superfluid's chirality arises from the condensation of atoms in the $p_+$ orbitals on the $B$ sublattice, which effectively creates a vortex lattice. Thus we expect a finite loop current correlation when the loops encircle the vortex cores and a vanishing correlation when they do not (see Fig.~\ref{fig3}). Indeed, using the ground state wave function in Eq.~(\ref{cwf}) we find that $\la \hat C_s \ra =- 3\sqrt{3}N(\cos{\xi} )^2$ and $\la\hat C_p^{\alpha\alpha'} \ra =0$, which leads to
\begin{align}
 \sigma_H(\omega)  = -\frac{\sqrt{3}\rho_u |t|^2}{\hbar^3\omega^2} (\cos \xi)^2
  \label{asym}
\end{align}
for $\hbar \omega \gg |t|$. This result is plotted in Fig.~\ref{fig2}(b) and agrees well with the exact calculation in the high frequency limit.

\begin{figure}[tbp]
\includegraphics[width=8.7cm]{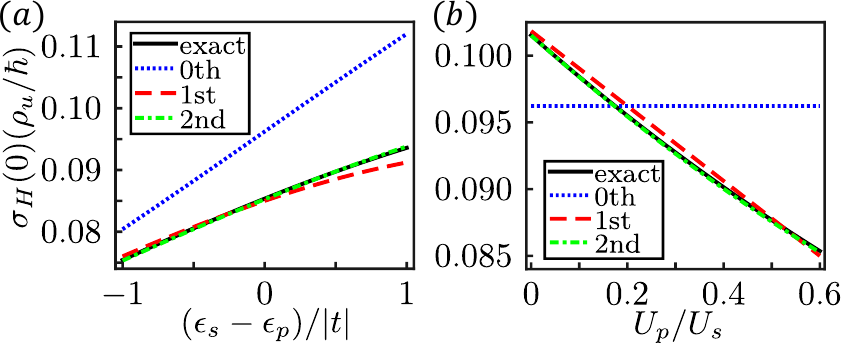}
\caption{\label{fig4} Perturbative analysis of the effect of atomic interactions on the dc Hall conductivity. Here results with second order corrections are included to show that the perturbative expansion converges very fast. (a) dc Hall conductivity as a function of the orbital energy differences. Here $\rho_uU_s =0.7|t| $ and $\rho_u U_p = 0.42|t|$ (b) dc Hall conductivity as a function of $U_p/U_s$. Here $\epsilon_s = \epsilon_p$ and  $\rho_uU_s =0.7 |t|$. }
\end{figure}

{\it dc Hall response and Berry curvature.}---At low frequencies, we first note that $\sigma_H(\omega)$ is an even function of $\omega$, implying that $\lim_{\omega\rightarrow 0}\sigma_H(\omega) = \sigma_H(0) + O(\omega^2)$. This explains the small deviations of $\sigma_H(\omega) $ from $\sigma_H(0)$ for a significant range of frequency. Next we explore the connections between the dc Hall conductivity $\sigma_H(0)$ and the Berry curvatures of the non-interacting and the Bogoliubov excitation bands. The Berry curvature of the latter is given by~\cite{2013Shindou,Engelhardt2015}
\begin{align}
\Omega_n(\bk) = i\epsilon_{ij}\sum_{m\neq n}\frac{V_n^\dag \pa_i \calH_B V_m \tau_{z,mm}V_m^\dag \pa_j \calH_B V_n}{(E_n-E_m)^2},
\label{Berry}
\end{align}
where $\epsilon_{ij}$ is the Levi-Civita symbol and repeated $i,j$ indices are summed over. We note that the $\tau_z$ matrix here arises from the normalization of the Bogliubov amplitudes~\cite{SM}. Furthermore, as the Berry curvature is a purely geometrical property of the Bogoliubov equation, the summation over the band index in Eq.~(\ref{Berry}) includes both the positive ($n=0,1,2$) and negative ($n=\bar 0,\bar 1, \bar 2$) branches,
which are related to each other by $E_n(\bk) = -E_{\bar n}(2\bK-\bk)$.  Since $n=0,\bk=\bK$ and $n=\bar 0,\bk=\bK$ denote the same ground state,  the dispersions $E_0(\bk)$ and $E_{\bar 0}(\bk)$ are joined at $\bk = \bK$, giving rise to a band touching point in the spectrum of $\calH_B$.  As a result the formula in Eq.~(\ref{Berry}) defines the Berry curvature of the Bogoliubov bands for all momenta except for the point of $n=0$ and $\bk = \bK$, which has no well-defined Berry curvature~\cite{2015Furukawa}. Nevertheless, the quantity $\Omega_0(\bK)$ as given by Eq.~(\ref{Berry}) is finite~\footnote{We note that in the expression for $\Omega_0(\bK)$ the summation over the band index excludes $\bar 0$ since $n=0,\bk=\bK$ and $n=\bar 0,\bk=\bK$ refer to the same ground state.} and in fact determines the low frequency Hall conductivity of the chiral superfluid. Comparing Eq.~(\ref{Hall}) and Eq.~({\ref{Berry}}) we immediately arrive at 
\begin{align}
\sigma_H(0) =\frac{N}{\hbar\calA} \Omega_0(\bK).
\end{align}
Even though $\Omega_0(\bK)$ no longer has the interpretation of the Berry curvature of the Bogoliubov band, its existence has the origin in the Berry curvature of the non-interacting band. Namely we find  $\Omega_0(\bK) =  \Omega_0^{(0)}(\bK) + O(U_s,U_p)$ in the limit of small atomic interactions, where $\Omega^{(0)}_n(\bk)$ is the Berry curvature of the non-interacting bands. The correction due to atomic interactions can be obtained perturbatively in terms of the scattering length~\cite{SM}. For experimentally relevant parameters this correction is significant and is well captured by a first order calculation, as shown in Fig.~\ref{fig4}(a) for a fixed scattering length. Such an agreement is also found for  a range of different on-site interaction strengths. This is best illustrated by the case of $\epsilon_s=\epsilon_p$, for which the first order result is given by
\begin{align}
\sigma_H(0) = \frac{N}{\hbar\calA } \left [ \Omega_0^{(0)}(\bK) -\frac{ \rho_u a^2 }{144|t|}(5U_p-U_s) \right ]. 
\end{align}
As shown in Fig.~\ref{fig4}(b), this result again provides an excellent account for the interaction induced correction of the dc Hall conductivity. 

Although the dc AHE discussed here is related to the Berry curvature, several important differences distinguish it from that found in electronic materials.  First, the finite Berry curvatures of the non-interacting bands here result from the broken inversion symmetry of the BN lattice potential rather than from the spin-orbit coupling. Secondly, quantum statistics plays a consequential role.  Because of the Bose condensation the superfluid has a finite dc Hall conductivity as long as the  non-interacting state corresponding to the condensation mode has a finite Berry curvature. In contrast, a normal Fermi gas in the same BN lattice would have a vanishing dc Hall conductivity, due to the Fermi statistics and the fact that $\Omega_n^{(0)}(-\bk) =- \Omega_n^{(0)}(\bk)$. Finally, the interaction induced correction to the Hall conductivity discussed above is unique to the chiral superfluid. 

{\it Experimental proposals.}--- Since atoms are charge neutral, transport properties of quantum gases cannot be measured with conventional condensed matter probes. We now outline relevant probes designed specifically for quantum gases and can be used to detect the AHE. The ac AHE can be detected by taking advantage of the fact that quantum gases are often trapped by an additional weak harmonic potential with frequency $\omega_{tr}$. A periodic displacement of the trapping potential along the $x$-direction at frequency $\omega$ and amplitude $d_x$  generates a force on the atoms. The resulting center of mass motion of the gas along the $y$-direction  can be measured by a quantum gas microscope~\cite{2016Kuhr} as $ R_y(\omega) \cos[\omega t - \phi_y(\omega)]$. The Hall conductivity can then be read out as
$
\sigma_H(\omega) = {\rm Im}[({N\omega}/{i\calA m \omega^2_{tr} d_x} )R_y (\omega) e^{i\phi_y(\omega)}]
$~\cite{2015Wu,SM} , 
where $m$ is the mass of the atom.
Such a probe has already been successfully implemented to measure the longitudinal  conductivity for a Fermi gas in optical lattice~\cite{2019Anderson}. However, this method is only suitable for measuring responses at frequency $\omega > \omega_{tr}$ and thus cannot access the dc conductivity. For the latter we can resort to the so-called dichroism probe~\cite{2017Tran,2019Asteria,2020Midtgaard}. More specifically, one considers a circular drive of the form 
$
V_\pm(\br) = 2 \mathcal E \left(x\cos{\bar\omega t} \pm  y \sin{\bar \omega t }\right)$, which induces a certain excitation rate  $\Gamma_\pm(\bar \omega)$ out of the ground state. The observable of interest is provided by the differential integrated rate  $\Delta\Gamma=\frac12\int_0^\infty\!d\bar \omega[\Gamma_{+}(\bar\omega)-\Gamma_{-}(\bar\omega)]$, which is related to the dc Hall conductivity via
$
\sigma_{H}(0)  =\Delta\Gamma /(2 \pi \calA\mathcal{E}^2)
$~\cite{SM}. This probe has also been successfully applied to measure the Chern number in a cold atomic realization of the Haldane model~\cite{2019Asteria}. 

Lastly, we comment on the effect of finite temperature on the experimental detection of the AHE. Our prediction of the AHE  is based on the assumption that the superfluid contains a significant condensate.  For trapped quasi-2D Bose gases, the superfluid will remain a true condensate for low temperatures and crosses over into the quasi-condensate regime as the temperature increases, before eventually undergoing the Berezinskii-Kosterlitz-Thouless (BKT) transition~\cite{Petrov2000}. Thus we expect that the AHE is detectable for temperatures much lower than the BKT transition temperature.

{\it Conclusions.}---In the context of atomic gases, the AHE refers to the phenomenon where the application of a uniform force induces a transverse current, in the absence of an artificial magnetic field or the rotation of any external trapping potential. We have shown that such an AHE exists in a recently realized bosonic chiral superfluid at zero temperature.  The causes for the AHE have been analyzed in both the low and high frequency limits. We find that the chirality of the system dictates the behavior of the high frequency response while finite Berry curvatures of the non-interacting bands underpin the dc Hall response. We point out experimental methods, already available and tested in other quantum gas systems, for the detection of this effect. Our findings are of immediate experimental interests and may spur further studies on bosonic topological matters.

\textit{Acknowledgement}.  We are grateful to Georg Bruun and Georg Engelhardt  for valuable discussions and for proof reading the manuscript. This work is supported by NSFC (Grant No.~11974161 and Grant No.~U1801661), Shenzhen Science and Technology Program (Grant No.~KQTD20200820113010023), the Key-Area Research and Development Program of Guangdong Province (Grant No.~2019B030330001), the National Key R\&D Program of China (Grant No.~2018YFA0307200), and a grant from Guangdong province (Grant No.~2019ZT08X324).


\begin{thebibliography}{55}%
\makeatletter
\providecommand \@ifxundefined [1]{%
 \@ifx{#1\undefined}
}%
\providecommand \@ifnum [1]{%
 \ifnum #1\expandafter \@firstoftwo
 \else \expandafter \@secondoftwo
 \fi
}%
\providecommand \@ifx [1]{%
 \ifx #1\expandafter \@firstoftwo
 \else \expandafter \@secondoftwo
 \fi
}%
\providecommand \natexlab [1]{#1}%
\providecommand \enquote  [1]{``#1''}%
\providecommand \bibnamefont  [1]{#1}%
\providecommand \bibfnamefont [1]{#1}%
\providecommand \citenamefont [1]{#1}%
\providecommand \href@noop [0]{\@secondoftwo}%
\providecommand \href [0]{\begingroup \@sanitize@url \@href}%
\providecommand \@href[1]{\@@startlink{#1}\@@href}%
\providecommand \@@href[1]{\endgroup#1\@@endlink}%
\providecommand \@sanitize@url [0]{\catcode `\\12\catcode `\$12\catcode
  `\&12\catcode `\#12\catcode `\^12\catcode `\_12\catcode `\%12\relax}%
\providecommand \@@startlink[1]{}%
\providecommand \@@endlink[0]{}%
\providecommand \url  [0]{\begingroup\@sanitize@url \@url }%
\providecommand \@url [1]{\endgroup\@href {#1}{\urlprefix }}%
\providecommand \urlprefix  [0]{URL }%
\providecommand \Eprint [0]{\href }%
\providecommand \doibase [0]{http://dx.doi.org/}%
\providecommand \selectlanguage [0]{\@gobble}%
\providecommand \bibinfo  [0]{\@secondoftwo}%
\providecommand \bibfield  [0]{\@secondoftwo}%
\providecommand \translation [1]{[#1]}%
\providecommand \BibitemOpen [0]{}%
\providecommand \bibitemStop [0]{}%
\providecommand \bibitemNoStop [0]{.\EOS\space}%
\providecommand \EOS [0]{\spacefactor3000\relax}%
\providecommand \BibitemShut  [1]{\csname bibitem#1\endcsname}%
\let\auto@bib@innerbib\@empty
\bibitem [{\citenamefont {Bloch}\ \emph {et~al.}(2012)\citenamefont {Bloch},
  \citenamefont {Dalibard},\ and\ \citenamefont {Nascimb{\`e}ne}}]{2012Bloch}%
  \BibitemOpen
  \bibfield  {author} {\bibinfo {author} {\bibfnamefont {I.}~\bibnamefont
  {Bloch}}, \bibinfo {author} {\bibfnamefont {J.}~\bibnamefont {Dalibard}}, \
  and\ \bibinfo {author} {\bibfnamefont {S.}~\bibnamefont {Nascimb{\`e}ne}},\
  }\href {\doibase 10.1038/nphys2259} {\bibfield  {journal} {\bibinfo
  {journal} {Nature Physics}\ }\textbf {\bibinfo {volume} {8}},\ \bibinfo
  {pages} {267} (\bibinfo {year} {2012})}\BibitemShut {NoStop}%
\bibitem [{\citenamefont {Georgescu}\ \emph {et~al.}(2014)\citenamefont
  {Georgescu}, \citenamefont {Ashhab},\ and\ \citenamefont
  {Nori}}]{2014Georgescu}%
  \BibitemOpen
  \bibfield  {author} {\bibinfo {author} {\bibfnamefont {I.~M.}\ \bibnamefont
  {Georgescu}}, \bibinfo {author} {\bibfnamefont {S.}~\bibnamefont {Ashhab}}, \
  and\ \bibinfo {author} {\bibfnamefont {F.}~\bibnamefont {Nori}},\ }\href
  {\doibase 10.1103/RevModPhys.86.153} {\bibfield  {journal} {\bibinfo
  {journal} {Rev. Mod. Phys.}\ }\textbf {\bibinfo {volume} {86}},\ \bibinfo
  {pages} {153} (\bibinfo {year} {2014})}\BibitemShut {NoStop}%
\bibitem [{\citenamefont {Zhai}(2021)}]{zhai_2021}%
  \BibitemOpen
  \bibfield  {author} {\bibinfo {author} {\bibfnamefont {H.}~\bibnamefont
  {Zhai}},\ }\href {\doibase 10.1017/9781108595216} {\emph {\bibinfo {title}
  {Ultracold Atomic Physics}}}\ (\bibinfo  {publisher} {Cambridge University
  Press},\ \bibinfo {year} {2021})\BibitemShut {NoStop}%
\bibitem [{\citenamefont {Jaksch}\ \emph {et~al.}(1998)\citenamefont {Jaksch},
  \citenamefont {Bruder}, \citenamefont {Cirac}, \citenamefont {Gardiner},\
  and\ \citenamefont {Zoller}}]{1998Jaksch}%
  \BibitemOpen
  \bibfield  {author} {\bibinfo {author} {\bibfnamefont {D.}~\bibnamefont
  {Jaksch}}, \bibinfo {author} {\bibfnamefont {C.}~\bibnamefont {Bruder}},
  \bibinfo {author} {\bibfnamefont {J.~I.}\ \bibnamefont {Cirac}}, \bibinfo
  {author} {\bibfnamefont {C.~W.}\ \bibnamefont {Gardiner}}, \ and\ \bibinfo
  {author} {\bibfnamefont {P.}~\bibnamefont {Zoller}},\ }\href {\doibase
  10.1103/PhysRevLett.81.3108} {\bibfield  {journal} {\bibinfo  {journal}
  {Phys. Rev. Lett.}\ }\textbf {\bibinfo {volume} {81}},\ \bibinfo {pages}
  {3108} (\bibinfo {year} {1998})}\BibitemShut {NoStop}%
\bibitem [{\citenamefont {Bloch}\ \emph {et~al.}(2008)\citenamefont {Bloch},
  \citenamefont {Dalibard},\ and\ \citenamefont {Zwerger}}]{2008Bloch}%
  \BibitemOpen
  \bibfield  {author} {\bibinfo {author} {\bibfnamefont {I.}~\bibnamefont
  {Bloch}}, \bibinfo {author} {\bibfnamefont {J.}~\bibnamefont {Dalibard}}, \
  and\ \bibinfo {author} {\bibfnamefont {W.}~\bibnamefont {Zwerger}},\ }\href
  {\doibase 10.1103/RevModPhys.80.885} {\bibfield  {journal} {\bibinfo
  {journal} {Rev. Mod. Phys.}\ }\textbf {\bibinfo {volume} {80}},\ \bibinfo
  {pages} {885} (\bibinfo {year} {2008})}\BibitemShut {NoStop}%
\bibitem [{\citenamefont {Lin}\ \emph {et~al.}(2011)\citenamefont {Lin},
  \citenamefont {Jim{\'e}nez-Garc{\'\i}a},\ and\ \citenamefont
  {Spielman}}]{2011Lin}%
  \BibitemOpen
  \bibfield  {author} {\bibinfo {author} {\bibfnamefont {Y.~J.}\ \bibnamefont
  {Lin}}, \bibinfo {author} {\bibfnamefont {K.}~\bibnamefont
  {Jim{\'e}nez-Garc{\'\i}a}}, \ and\ \bibinfo {author} {\bibfnamefont {I.~B.}\
  \bibnamefont {Spielman}},\ }\href {\doibase 10.1038/nature09887} {\bibfield
  {journal} {\bibinfo  {journal} {Nature}\ }\textbf {\bibinfo {volume} {471}},\
  \bibinfo {pages} {83} (\bibinfo {year} {2011})}\BibitemShut {NoStop}%
\bibitem [{\citenamefont {Galitski}\ and\ \citenamefont
  {Spielman}(2013)}]{2013Galitski}%
  \BibitemOpen
  \bibfield  {author} {\bibinfo {author} {\bibfnamefont {V.}~\bibnamefont
  {Galitski}}\ and\ \bibinfo {author} {\bibfnamefont {I.~B.}\ \bibnamefont
  {Spielman}},\ }\href {\doibase 10.1038/nature11841} {\bibfield  {journal}
  {\bibinfo  {journal} {Nature}\ }\textbf {\bibinfo {volume} {494}},\ \bibinfo
  {pages} {49} (\bibinfo {year} {2013})}\BibitemShut {NoStop}%
\bibitem [{\citenamefont {Zhai}(2015)}]{Zhai_2015}%
  \BibitemOpen
  \bibfield  {author} {\bibinfo {author} {\bibfnamefont {H.}~\bibnamefont
  {Zhai}},\ }\href {\doibase 10.1088/0034-4885/78/2/026001} {\bibfield
  {journal} {\bibinfo  {journal} {Reports on Progress in Physics}\ }\textbf
  {\bibinfo {volume} {78}},\ \bibinfo {pages} {026001} (\bibinfo {year}
  {2015})}\BibitemShut {NoStop}%
\bibitem [{\citenamefont {Dalibard}\ \emph {et~al.}(2011)\citenamefont
  {Dalibard}, \citenamefont {Gerbier}, \citenamefont
  {Juzeli\ifmmode~\bar{u}\else \={u}\fi{}nas},\ and\ \citenamefont
  {\"Ohberg}}]{2011Dalibard}%
  \BibitemOpen
  \bibfield  {author} {\bibinfo {author} {\bibfnamefont {J.}~\bibnamefont
  {Dalibard}}, \bibinfo {author} {\bibfnamefont {F.}~\bibnamefont {Gerbier}},
  \bibinfo {author} {\bibfnamefont {G.}~\bibnamefont
  {Juzeli\ifmmode~\bar{u}\else \={u}\fi{}nas}}, \ and\ \bibinfo {author}
  {\bibfnamefont {P.}~\bibnamefont {\"Ohberg}},\ }\href {\doibase
  10.1103/RevModPhys.83.1523} {\bibfield  {journal} {\bibinfo  {journal} {Rev.
  Mod. Phys.}\ }\textbf {\bibinfo {volume} {83}},\ \bibinfo {pages} {1523}
  (\bibinfo {year} {2011})}\BibitemShut {NoStop}%
\bibitem [{\citenamefont {LeBlanc}\ \emph {et~al.}(2012)\citenamefont
  {LeBlanc}, \citenamefont {Jiménez-García}, \citenamefont {Williams},
  \citenamefont {Beeler}, \citenamefont {Perry}, \citenamefont {Phillips},\
  and\ \citenamefont {Spielman}}]{2012Leblanc}%
  \BibitemOpen
  \bibfield  {author} {\bibinfo {author} {\bibfnamefont {L.~J.}\ \bibnamefont
  {LeBlanc}}, \bibinfo {author} {\bibfnamefont {K.}~\bibnamefont
  {Jiménez-García}}, \bibinfo {author} {\bibfnamefont {R.~A.}\ \bibnamefont
  {Williams}}, \bibinfo {author} {\bibfnamefont {M.~C.}\ \bibnamefont
  {Beeler}}, \bibinfo {author} {\bibfnamefont {A.~R.}\ \bibnamefont {Perry}},
  \bibinfo {author} {\bibfnamefont {W.~D.}\ \bibnamefont {Phillips}}, \ and\
  \bibinfo {author} {\bibfnamefont {I.~B.}\ \bibnamefont {Spielman}},\ }\href
  {\doibase 10.1073/pnas.1202579109} {\bibfield  {journal} {\bibinfo  {journal}
  {Proceedings of the National Academy of Sciences}\ }\textbf {\bibinfo
  {volume} {109}},\ \bibinfo {pages} {10811} (\bibinfo {year}
  {2012})}\BibitemShut {NoStop}%
\bibitem [{\citenamefont {Tokura}\ and\ \citenamefont
  {Nagaosa}(2000)}]{2000Tokura}%
  \BibitemOpen
  \bibfield  {author} {\bibinfo {author} {\bibfnamefont {Y.}~\bibnamefont
  {Tokura}}\ and\ \bibinfo {author} {\bibfnamefont {N.}~\bibnamefont
  {Nagaosa}},\ }\href {\doibase 10.1126/science.288.5465.462} {\bibfield
  {journal} {\bibinfo  {journal} {Science}\ }\textbf {\bibinfo {volume}
  {288}},\ \bibinfo {pages} {462} (\bibinfo {year} {2000})}\BibitemShut
  {NoStop}%
\bibitem [{\citenamefont {Kock}\ \emph {et~al.}(2016)\citenamefont {Kock},
  \citenamefont {Hippler}, \citenamefont {Ewerbeck},\ and\ \citenamefont
  {Hemmerich}}]{2016Kock}%
  \BibitemOpen
  \bibfield  {author} {\bibinfo {author} {\bibfnamefont {T.}~\bibnamefont
  {Kock}}, \bibinfo {author} {\bibfnamefont {C.}~\bibnamefont {Hippler}},
  \bibinfo {author} {\bibfnamefont {A.}~\bibnamefont {Ewerbeck}}, \ and\
  \bibinfo {author} {\bibfnamefont {A.}~\bibnamefont {Hemmerich}},\ }\href
  {\doibase 10.1088/0953-4075/49/4/042001} {\bibfield  {journal} {\bibinfo
  {journal} {Journal of Physics B: Atomic, Molecular and Optical Physics}\
  }\textbf {\bibinfo {volume} {49}},\ \bibinfo {pages} {042001} (\bibinfo
  {year} {2016})}\BibitemShut {NoStop}%
\bibitem [{\citenamefont {Li}\ and\ \citenamefont
  {Liu}(2016)}]{2016XiaopengLi}%
  \BibitemOpen
  \bibfield  {author} {\bibinfo {author} {\bibfnamefont {X.}~\bibnamefont
  {Li}}\ and\ \bibinfo {author} {\bibfnamefont {W.~V.}\ \bibnamefont {Liu}},\
  }\href {\doibase 10.1088/0034-4885/79/11/116401} {\bibfield  {journal}
  {\bibinfo  {journal} {Reports on Progress in Physics}\ }\textbf {\bibinfo
  {volume} {79}},\ \bibinfo {pages} {116401} (\bibinfo {year}
  {2016})}\BibitemShut {NoStop}%
\bibitem [{\citenamefont {M\"uller}\ \emph {et~al.}(2007)\citenamefont
  {M\"uller}, \citenamefont {F\"olling}, \citenamefont {Widera},\ and\
  \citenamefont {Bloch}}]{2007Muller}%
  \BibitemOpen
  \bibfield  {author} {\bibinfo {author} {\bibfnamefont {T.}~\bibnamefont
  {M\"uller}}, \bibinfo {author} {\bibfnamefont {S.}~\bibnamefont {F\"olling}},
  \bibinfo {author} {\bibfnamefont {A.}~\bibnamefont {Widera}}, \ and\ \bibinfo
  {author} {\bibfnamefont {I.}~\bibnamefont {Bloch}},\ }\href {\doibase
  10.1103/PhysRevLett.99.200405} {\bibfield  {journal} {\bibinfo  {journal}
  {Phys. Rev. Lett.}\ }\textbf {\bibinfo {volume} {99}},\ \bibinfo {pages}
  {200405} (\bibinfo {year} {2007})}\BibitemShut {NoStop}%
\bibitem [{\citenamefont {Wirth}\ \emph {et~al.}(2011)\citenamefont {Wirth},
  \citenamefont {{\"O}lschl{\"a}ger},\ and\ \citenamefont
  {Hemmerich}}]{2010Wirth}%
  \BibitemOpen
  \bibfield  {author} {\bibinfo {author} {\bibfnamefont {G.}~\bibnamefont
  {Wirth}}, \bibinfo {author} {\bibfnamefont {M.}~\bibnamefont
  {{\"O}lschl{\"a}ger}}, \ and\ \bibinfo {author} {\bibfnamefont
  {A.}~\bibnamefont {Hemmerich}},\ }\href {\doibase 10.1038/nphys1857}
  {\bibfield  {journal} {\bibinfo  {journal} {Nature Physics}\ }\textbf
  {\bibinfo {volume} {7}},\ \bibinfo {pages} {147} (\bibinfo {year}
  {2011})}\BibitemShut {NoStop}%
\bibitem [{\citenamefont {Soltan-Panahi}\ \emph {et~al.}(2012)\citenamefont
  {Soltan-Panahi}, \citenamefont {L{\"u}hmann}, \citenamefont {Struck},
  \citenamefont {Windpassinger},\ and\ \citenamefont {Sengstock}}]{2012Soltan}%
  \BibitemOpen
  \bibfield  {author} {\bibinfo {author} {\bibfnamefont {P.}~\bibnamefont
  {Soltan-Panahi}}, \bibinfo {author} {\bibfnamefont {D.-S.}\ \bibnamefont
  {L{\"u}hmann}}, \bibinfo {author} {\bibfnamefont {J.}~\bibnamefont {Struck}},
  \bibinfo {author} {\bibfnamefont {P.}~\bibnamefont {Windpassinger}}, \ and\
  \bibinfo {author} {\bibfnamefont {K.}~\bibnamefont {Sengstock}},\ }\href
  {\doibase 10.1038/nphys2128} {\bibfield  {journal} {\bibinfo  {journal}
  {Nature Physics}\ }\textbf {\bibinfo {volume} {8}},\ \bibinfo {pages} {71}
  (\bibinfo {year} {2012})}\BibitemShut {NoStop}%
\bibitem [{\citenamefont {Ölschläger}\ \emph {et~al.}(2013)\citenamefont
  {Ölschläger}, \citenamefont {Kock}, \citenamefont {Wirth}, \citenamefont
  {Ewerbeck}, \citenamefont {Smith},\ and\ \citenamefont
  {Hemmerich}}]{2013Olschlager}%
  \BibitemOpen
  \bibfield  {author} {\bibinfo {author} {\bibfnamefont {M.}~\bibnamefont
  {Ölschläger}}, \bibinfo {author} {\bibfnamefont {T.}~\bibnamefont {Kock}},
  \bibinfo {author} {\bibfnamefont {G.}~\bibnamefont {Wirth}}, \bibinfo
  {author} {\bibfnamefont {A.}~\bibnamefont {Ewerbeck}}, \bibinfo {author}
  {\bibfnamefont {C.~M.}\ \bibnamefont {Smith}}, \ and\ \bibinfo {author}
  {\bibfnamefont {A.}~\bibnamefont {Hemmerich}},\ }\href {\doibase
  10.1088/1367-2630/15/8/083041} {\bibfield  {journal} {\bibinfo  {journal}
  {New Journal of Physics}\ }\textbf {\bibinfo {volume} {15}},\ \bibinfo
  {pages} {083041} (\bibinfo {year} {2013})}\BibitemShut {NoStop}%
\bibitem [{\citenamefont {Kock}\ \emph {et~al.}(2015)\citenamefont {Kock},
  \citenamefont {\"Olschl\"ager}, \citenamefont {Ewerbeck}, \citenamefont
  {Huang}, \citenamefont {Mathey},\ and\ \citenamefont {Hemmerich}}]{2015Kock}%
  \BibitemOpen
  \bibfield  {author} {\bibinfo {author} {\bibfnamefont {T.}~\bibnamefont
  {Kock}}, \bibinfo {author} {\bibfnamefont {M.}~\bibnamefont
  {\"Olschl\"ager}}, \bibinfo {author} {\bibfnamefont {A.}~\bibnamefont
  {Ewerbeck}}, \bibinfo {author} {\bibfnamefont {W.-M.}\ \bibnamefont {Huang}},
  \bibinfo {author} {\bibfnamefont {L.}~\bibnamefont {Mathey}}, \ and\ \bibinfo
  {author} {\bibfnamefont {A.}~\bibnamefont {Hemmerich}},\ }\href {\doibase
  10.1103/PhysRevLett.114.115301} {\bibfield  {journal} {\bibinfo  {journal}
  {Phys. Rev. Lett.}\ }\textbf {\bibinfo {volume} {114}},\ \bibinfo {pages}
  {115301} (\bibinfo {year} {2015})}\BibitemShut {NoStop}%
\bibitem [{\citenamefont {Weinberg}\ \emph {et~al.}(2016)\citenamefont
  {Weinberg}, \citenamefont {Staarmann}, \citenamefont {Ölschläger},
  \citenamefont {Simonet},\ and\ \citenamefont {Sengstock}}]{2016Sengstock}%
  \BibitemOpen
  \bibfield  {author} {\bibinfo {author} {\bibfnamefont {M.}~\bibnamefont
  {Weinberg}}, \bibinfo {author} {\bibfnamefont {C.}~\bibnamefont {Staarmann}},
  \bibinfo {author} {\bibfnamefont {C.}~\bibnamefont {Ölschläger}}, \bibinfo
  {author} {\bibfnamefont {J.}~\bibnamefont {Simonet}}, \ and\ \bibinfo
  {author} {\bibfnamefont {K.}~\bibnamefont {Sengstock}},\ }\href {\doibase
  10.1088/2053-1583/3/2/024005} {\bibfield  {journal} {\bibinfo  {journal} {2D
  Materials}\ }\textbf {\bibinfo {volume} {3}},\ \bibinfo {pages} {024005}
  (\bibinfo {year} {2016})}\BibitemShut {NoStop}%
\bibitem [{\citenamefont {Hachmann}\ \emph {et~al.}(2021)\citenamefont
  {Hachmann}, \citenamefont {Kiefer}, \citenamefont {Riebesehl}, \citenamefont
  {Eichberger},\ and\ \citenamefont {Hemmerich}}]{2021Hachmann}%
  \BibitemOpen
  \bibfield  {author} {\bibinfo {author} {\bibfnamefont {M.}~\bibnamefont
  {Hachmann}}, \bibinfo {author} {\bibfnamefont {Y.}~\bibnamefont {Kiefer}},
  \bibinfo {author} {\bibfnamefont {J.}~\bibnamefont {Riebesehl}}, \bibinfo
  {author} {\bibfnamefont {R.}~\bibnamefont {Eichberger}}, \ and\ \bibinfo
  {author} {\bibfnamefont {A.}~\bibnamefont {Hemmerich}},\ }\href {\doibase
  10.1103/PhysRevLett.127.033201} {\bibfield  {journal} {\bibinfo  {journal}
  {Phys. Rev. Lett.}\ }\textbf {\bibinfo {volume} {127}},\ \bibinfo {pages}
  {033201} (\bibinfo {year} {2021})}\BibitemShut {NoStop}%
\bibitem [{\citenamefont {Vargas}\ \emph {et~al.}(2021)\citenamefont {Vargas},
  \citenamefont {Nuske}, \citenamefont {Eichberger}, \citenamefont {Hippler},
  \citenamefont {Mathey},\ and\ \citenamefont {Hemmerich}}]{2021Vargas}%
  \BibitemOpen
  \bibfield  {author} {\bibinfo {author} {\bibfnamefont {J.}~\bibnamefont
  {Vargas}}, \bibinfo {author} {\bibfnamefont {M.}~\bibnamefont {Nuske}},
  \bibinfo {author} {\bibfnamefont {R.}~\bibnamefont {Eichberger}}, \bibinfo
  {author} {\bibfnamefont {C.}~\bibnamefont {Hippler}}, \bibinfo {author}
  {\bibfnamefont {L.}~\bibnamefont {Mathey}}, \ and\ \bibinfo {author}
  {\bibfnamefont {A.}~\bibnamefont {Hemmerich}},\ }\href {\doibase
  10.1103/PhysRevLett.126.200402} {\bibfield  {journal} {\bibinfo  {journal}
  {Phys. Rev. Lett.}\ }\textbf {\bibinfo {volume} {126}},\ \bibinfo {pages}
  {200402} (\bibinfo {year} {2021})}\BibitemShut {NoStop}%
\bibitem [{\citenamefont {Jin}\ \emph {et~al.}(2021)\citenamefont {Jin},
  \citenamefont {Zhang}, \citenamefont {Guo}, \citenamefont {Chen},
  \citenamefont {Zhou},\ and\ \citenamefont {Li}}]{2021Jin}%
  \BibitemOpen
  \bibfield  {author} {\bibinfo {author} {\bibfnamefont {S.}~\bibnamefont
  {Jin}}, \bibinfo {author} {\bibfnamefont {W.}~\bibnamefont {Zhang}}, \bibinfo
  {author} {\bibfnamefont {X.}~\bibnamefont {Guo}}, \bibinfo {author}
  {\bibfnamefont {X.}~\bibnamefont {Chen}}, \bibinfo {author} {\bibfnamefont
  {X.}~\bibnamefont {Zhou}}, \ and\ \bibinfo {author} {\bibfnamefont
  {X.}~\bibnamefont {Li}},\ }\href {\doibase 10.1103/PhysRevLett.126.035301}
  {\bibfield  {journal} {\bibinfo  {journal} {Phys. Rev. Lett.}\ }\textbf
  {\bibinfo {volume} {126}},\ \bibinfo {pages} {035301} (\bibinfo {year}
  {2021})}\BibitemShut {NoStop}%
\bibitem [{\citenamefont {Wang}\ \emph {et~al.}(2021)\citenamefont {Wang},
  \citenamefont {Luo}, \citenamefont {Liu}, \citenamefont {Liu}, \citenamefont
  {Hemmerich},\ and\ \citenamefont {Xu}}]{2021Wang}%
  \BibitemOpen
  \bibfield  {author} {\bibinfo {author} {\bibfnamefont {X.-Q.}\ \bibnamefont
  {Wang}}, \bibinfo {author} {\bibfnamefont {G.-Q.}\ \bibnamefont {Luo}},
  \bibinfo {author} {\bibfnamefont {J.-Y.}\ \bibnamefont {Liu}}, \bibinfo
  {author} {\bibfnamefont {W.~V.}\ \bibnamefont {Liu}}, \bibinfo {author}
  {\bibfnamefont {A.}~\bibnamefont {Hemmerich}}, \ and\ \bibinfo {author}
  {\bibfnamefont {Z.-F.}\ \bibnamefont {Xu}},\ }\href {\doibase
  10.1038/s41586-021-03702-0} {\bibfield  {journal} {\bibinfo  {journal}
  {Nature}\ }\textbf {\bibinfo {volume} {596}},\ \bibinfo {pages} {227}
  (\bibinfo {year} {2021})}\BibitemShut {NoStop}%
\bibitem [{\citenamefont {Walmsley}\ and\ \citenamefont
  {Golov}(2012)}]{2012Walmsley}%
  \BibitemOpen
  \bibfield  {author} {\bibinfo {author} {\bibfnamefont {P.~M.}\ \bibnamefont
  {Walmsley}}\ and\ \bibinfo {author} {\bibfnamefont {A.~I.}\ \bibnamefont
  {Golov}},\ }\href {\doibase 10.1103/PhysRevLett.109.215301} {\bibfield
  {journal} {\bibinfo  {journal} {Phys. Rev. Lett.}\ }\textbf {\bibinfo
  {volume} {109}},\ \bibinfo {pages} {215301} (\bibinfo {year}
  {2012})}\BibitemShut {NoStop}%
\bibitem [{\citenamefont {Ikegami}\ \emph {et~al.}(2013)\citenamefont
  {Ikegami}, \citenamefont {Tsutsumi},\ and\ \citenamefont
  {Kono}}]{2013Ikegami}%
  \BibitemOpen
  \bibfield  {author} {\bibinfo {author} {\bibfnamefont {H.}~\bibnamefont
  {Ikegami}}, \bibinfo {author} {\bibfnamefont {Y.}~\bibnamefont {Tsutsumi}}, \
  and\ \bibinfo {author} {\bibfnamefont {K.}~\bibnamefont {Kono}},\ }\href
  {\doibase 10.1126/science.1236509} {\bibfield  {journal} {\bibinfo  {journal}
  {Science}\ }\textbf {\bibinfo {volume} {341}},\ \bibinfo {pages} {59}
  (\bibinfo {year} {2013})}\BibitemShut {NoStop}%
\bibitem [{\citenamefont {Bernevig}\ and\ \citenamefont
  {Hughes}(2013)}]{2013Bernevig}%
  \BibitemOpen
  \bibfield  {author} {\bibinfo {author} {\bibfnamefont {B.~A.}\ \bibnamefont
  {Bernevig}}\ and\ \bibinfo {author} {\bibfnamefont {T.~L.}\ \bibnamefont
  {Hughes}},\ }\href@noop {} {\emph {\bibinfo {title} {Topological Insulators
  and Topological Superconductors}}}\ (\bibinfo  {publisher} {Princeton
  University Press, Princeton, NJ},\ \bibinfo {year} {2013})\BibitemShut
  {NoStop}%
\bibitem [{\citenamefont {Thouless}\ \emph {et~al.}(1982)\citenamefont
  {Thouless}, \citenamefont {Kohmoto}, \citenamefont {Nightingale},\ and\
  \citenamefont {den Nijs}}]{1982TKNN}%
  \BibitemOpen
  \bibfield  {author} {\bibinfo {author} {\bibfnamefont {D.~J.}\ \bibnamefont
  {Thouless}}, \bibinfo {author} {\bibfnamefont {M.}~\bibnamefont {Kohmoto}},
  \bibinfo {author} {\bibfnamefont {M.~P.}\ \bibnamefont {Nightingale}}, \ and\
  \bibinfo {author} {\bibfnamefont {M.}~\bibnamefont {den Nijs}},\ }\href
  {\doibase 10.1103/PhysRevLett.49.405} {\bibfield  {journal} {\bibinfo
  {journal} {Phys. Rev. Lett.}\ }\textbf {\bibinfo {volume} {49}},\ \bibinfo
  {pages} {405} (\bibinfo {year} {1982})}\BibitemShut {NoStop}%
\bibitem [{\citenamefont {Haldane}(1988)}]{1988Haldane}%
  \BibitemOpen
  \bibfield  {author} {\bibinfo {author} {\bibfnamefont {F.~D.~M.}\
  \bibnamefont {Haldane}},\ }\href {\doibase 10.1103/PhysRevLett.61.2015}
  {\bibfield  {journal} {\bibinfo  {journal} {Phys. Rev. Lett.}\ }\textbf
  {\bibinfo {volume} {61}},\ \bibinfo {pages} {2015} (\bibinfo {year}
  {1988})}\BibitemShut {NoStop}%
\bibitem [{\citenamefont {Liu}\ \emph {et~al.}(2016)\citenamefont {Liu},
  \citenamefont {Zhang},\ and\ \citenamefont {Qi}}]{2016Liu}%
  \BibitemOpen
  \bibfield  {author} {\bibinfo {author} {\bibfnamefont {C.-X.}\ \bibnamefont
  {Liu}}, \bibinfo {author} {\bibfnamefont {S.-C.}\ \bibnamefont {Zhang}}, \
  and\ \bibinfo {author} {\bibfnamefont {X.-L.}\ \bibnamefont {Qi}},\ }\href
  {\doibase 10.1146/annurev-conmatphys-031115-011417} {\bibfield  {journal}
  {\bibinfo  {journal} {Annual Review of Condensed Matter Physics}\ }\textbf
  {\bibinfo {volume} {7}},\ \bibinfo {pages} {301} (\bibinfo {year}
  {2016})}\BibitemShut {NoStop}%
\bibitem [{\citenamefont {Nagaosa}\ \emph {et~al.}(2010)\citenamefont
  {Nagaosa}, \citenamefont {Sinova}, \citenamefont {Onoda}, \citenamefont
  {MacDonald},\ and\ \citenamefont {Ong}}]{2010Nagaosa}%
  \BibitemOpen
  \bibfield  {author} {\bibinfo {author} {\bibfnamefont {N.}~\bibnamefont
  {Nagaosa}}, \bibinfo {author} {\bibfnamefont {J.}~\bibnamefont {Sinova}},
  \bibinfo {author} {\bibfnamefont {S.}~\bibnamefont {Onoda}}, \bibinfo
  {author} {\bibfnamefont {A.~H.}\ \bibnamefont {MacDonald}}, \ and\ \bibinfo
  {author} {\bibfnamefont {N.~P.}\ \bibnamefont {Ong}},\ }\href {\doibase
  10.1103/RevModPhys.82.1539} {\bibfield  {journal} {\bibinfo  {journal} {Rev.
  Mod. Phys.}\ }\textbf {\bibinfo {volume} {82}},\ \bibinfo {pages} {1539}
  (\bibinfo {year} {2010})}\BibitemShut {NoStop}%
\bibitem [{\citenamefont {Xiao}\ \emph {et~al.}(2010)\citenamefont {Xiao},
  \citenamefont {Chang},\ and\ \citenamefont {Niu}}]{2010Xiao}%
  \BibitemOpen
  \bibfield  {author} {\bibinfo {author} {\bibfnamefont {D.}~\bibnamefont
  {Xiao}}, \bibinfo {author} {\bibfnamefont {M.-C.}\ \bibnamefont {Chang}}, \
  and\ \bibinfo {author} {\bibfnamefont {Q.}~\bibnamefont {Niu}},\ }\href
  {\doibase 10.1103/RevModPhys.82.1959} {\bibfield  {journal} {\bibinfo
  {journal} {Rev. Mod. Phys.}\ }\textbf {\bibinfo {volume} {82}},\ \bibinfo
  {pages} {1959} (\bibinfo {year} {2010})}\BibitemShut {NoStop}%
\bibitem [{\citenamefont {Dudarev}\ \emph {et~al.}(2004)\citenamefont
  {Dudarev}, \citenamefont {Diener}, \citenamefont {Carusotto},\ and\
  \citenamefont {Niu}}]{2004Dudarev}%
  \BibitemOpen
  \bibfield  {author} {\bibinfo {author} {\bibfnamefont {A.~M.}\ \bibnamefont
  {Dudarev}}, \bibinfo {author} {\bibfnamefont {R.~B.}\ \bibnamefont {Diener}},
  \bibinfo {author} {\bibfnamefont {I.}~\bibnamefont {Carusotto}}, \ and\
  \bibinfo {author} {\bibfnamefont {Q.}~\bibnamefont {Niu}},\ }\href {\doibase
  10.1103/PhysRevLett.92.153005} {\bibfield  {journal} {\bibinfo  {journal}
  {Phys. Rev. Lett.}\ }\textbf {\bibinfo {volume} {92}},\ \bibinfo {pages}
  {153005} (\bibinfo {year} {2004})}\BibitemShut {NoStop}%
\bibitem [{\citenamefont {Li}\ \emph {et~al.}(2015)\citenamefont {Li},
  \citenamefont {Sengupta}, \citenamefont {Batrouni}, \citenamefont
  {Miniatura},\ and\ \citenamefont {Gr\'emaud}}]{2015Li}%
  \BibitemOpen
  \bibfield  {author} {\bibinfo {author} {\bibfnamefont {Y.}~\bibnamefont
  {Li}}, \bibinfo {author} {\bibfnamefont {P.}~\bibnamefont {Sengupta}},
  \bibinfo {author} {\bibfnamefont {G.~G.}\ \bibnamefont {Batrouni}}, \bibinfo
  {author} {\bibfnamefont {C.}~\bibnamefont {Miniatura}}, \ and\ \bibinfo
  {author} {\bibfnamefont {B.}~\bibnamefont {Gr\'emaud}},\ }\href {\doibase
  10.1103/PhysRevA.92.043605} {\bibfield  {journal} {\bibinfo  {journal} {Phys.
  Rev. A}\ }\textbf {\bibinfo {volume} {92}},\ \bibinfo {pages} {043605}
  (\bibinfo {year} {2015})}\BibitemShut {NoStop}%
\bibitem [{\citenamefont {van~der Bijl}\ and\ \citenamefont
  {Duine}(2011)}]{2011Bijl}%
  \BibitemOpen
  \bibfield  {author} {\bibinfo {author} {\bibfnamefont {E.}~\bibnamefont
  {van~der Bijl}}\ and\ \bibinfo {author} {\bibfnamefont {R.~A.}\ \bibnamefont
  {Duine}},\ }\href {\doibase 10.1103/PhysRevLett.107.195302} {\bibfield
  {journal} {\bibinfo  {journal} {Phys. Rev. Lett.}\ }\textbf {\bibinfo
  {volume} {107}},\ \bibinfo {pages} {195302} (\bibinfo {year}
  {2011})}\BibitemShut {NoStop}%
\bibitem [{\citenamefont {Patucha}\ \emph {et~al.}(2018)\citenamefont
  {Patucha}, \citenamefont {Grygiel},\ and\ \citenamefont
  {Zaleski}}]{2018patucha}%
  \BibitemOpen
  \bibfield  {author} {\bibinfo {author} {\bibfnamefont {K.}~\bibnamefont
  {Patucha}}, \bibinfo {author} {\bibfnamefont {B.}~\bibnamefont {Grygiel}}, \
  and\ \bibinfo {author} {\bibfnamefont {T.~A.}\ \bibnamefont {Zaleski}},\
  }\href {\doibase 10.1103/PhysRevB.97.214522} {\bibfield  {journal} {\bibinfo
  {journal} {Phys. Rev. B}\ }\textbf {\bibinfo {volume} {97}},\ \bibinfo
  {pages} {214522} (\bibinfo {year} {2018})}\BibitemShut {NoStop}%
\bibitem [{\citenamefont {Taylor}\ and\ \citenamefont
  {Kallin}(2012)}]{2012Taylor}%
  \BibitemOpen
  \bibfield  {author} {\bibinfo {author} {\bibfnamefont {E.}~\bibnamefont
  {Taylor}}\ and\ \bibinfo {author} {\bibfnamefont {C.}~\bibnamefont
  {Kallin}},\ }\href {\doibase 10.1103/PhysRevLett.108.157001} {\bibfield
  {journal} {\bibinfo  {journal} {Phys. Rev. Lett.}\ }\textbf {\bibinfo
  {volume} {108}},\ \bibinfo {pages} {157001} (\bibinfo {year}
  {2012})}\BibitemShut {NoStop}%
\bibitem [{\citenamefont {Lutchyn}\ \emph {et~al.}(2009)\citenamefont
  {Lutchyn}, \citenamefont {Nagornykh},\ and\ \citenamefont
  {Yakovenko}}]{2009Lutchyn}%
  \BibitemOpen
  \bibfield  {author} {\bibinfo {author} {\bibfnamefont {R.~M.}\ \bibnamefont
  {Lutchyn}}, \bibinfo {author} {\bibfnamefont {P.}~\bibnamefont {Nagornykh}},
  \ and\ \bibinfo {author} {\bibfnamefont {V.~M.}\ \bibnamefont {Yakovenko}},\
  }\href {\doibase 10.1103/PhysRevB.80.104508} {\bibfield  {journal} {\bibinfo
  {journal} {Phys. Rev. B}\ }\textbf {\bibinfo {volume} {80}},\ \bibinfo
  {pages} {104508} (\bibinfo {year} {2009})}\BibitemShut {NoStop}%
\bibitem [{\citenamefont {Kallin}\ and\ \citenamefont
  {Berlinsky}(2016)}]{2016Kallin}%
  \BibitemOpen
  \bibfield  {author} {\bibinfo {author} {\bibfnamefont {C.}~\bibnamefont
  {Kallin}}\ and\ \bibinfo {author} {\bibfnamefont {J.}~\bibnamefont
  {Berlinsky}},\ }\href {\doibase 10.1088/0034-4885/79/5/054502} {\bibfield
  {journal} {\bibinfo  {journal} {Reports on Progress in Physics}\ }\textbf
  {\bibinfo {volume} {79}},\ \bibinfo {pages} {054502} (\bibinfo {year}
  {2016})}\BibitemShut {NoStop}%
\bibitem [{\citenamefont {Brydon}\ \emph {et~al.}(2019)\citenamefont {Brydon},
  \citenamefont {Abergel}, \citenamefont {Agterberg},\ and\ \citenamefont
  {Yakovenko}}]{2019Brydon}%
  \BibitemOpen
  \bibfield  {author} {\bibinfo {author} {\bibfnamefont {P.~M.~R.}\
  \bibnamefont {Brydon}}, \bibinfo {author} {\bibfnamefont {D.~S.~L.}\
  \bibnamefont {Abergel}}, \bibinfo {author} {\bibfnamefont {D.~F.}\
  \bibnamefont {Agterberg}}, \ and\ \bibinfo {author} {\bibfnamefont {V.~M.}\
  \bibnamefont {Yakovenko}},\ }\href {\doibase 10.1103/PhysRevX.9.031025}
  {\bibfield  {journal} {\bibinfo  {journal} {Phys. Rev. X}\ }\textbf {\bibinfo
  {volume} {9}},\ \bibinfo {pages} {031025} (\bibinfo {year}
  {2019})}\BibitemShut {NoStop}%
\bibitem [{\citenamefont {Denys}\ and\ \citenamefont
  {Brydon}(2021)}]{Denys2021}%
  \BibitemOpen
  \bibfield  {author} {\bibinfo {author} {\bibfnamefont {M.~D.~E.}\
  \bibnamefont {Denys}}\ and\ \bibinfo {author} {\bibfnamefont {P.~M.~R.}\
  \bibnamefont {Brydon}},\ }\href {\doibase 10.1103/PhysRevB.103.094503}
  {\bibfield  {journal} {\bibinfo  {journal} {Phys. Rev. B}\ }\textbf {\bibinfo
  {volume} {103}},\ \bibinfo {pages} {094503} (\bibinfo {year}
  {2021})}\BibitemShut {NoStop}%
\bibitem [{SM()}]{SM}%
  \BibitemOpen
  \href@noop {} {\bibinfo  {journal} {See Supplemental Material for more
  details on the multi-orbital Bose-Hubbard model, the superfluid ground state,
  the symmetry properties of the $\bK$ and $\bK'$ points, the loop currents,
  the perturbative treatment of the Hall conductivity, the Berry curvature of
  the Bogoliubov bands, and the experimental probes, which includes
  Ref.~\cite{Marzari2012}}\ }\BibitemShut {NoStop}%
  \bibitem [{\citenamefont {Marzari}\ \emph {et~al.}(2012)\citenamefont
  {Marzari}, \citenamefont {Mostofi}, \citenamefont {Yates}, \citenamefont
  {Souza},\ and\ \citenamefont {Vanderbilt}}]{Marzari2012}%
  \BibitemOpen
  \bibfield  {author} {\bibinfo {author} {\bibfnamefont {N.}~\bibnamefont
  {Marzari}}, \bibinfo {author} {\bibfnamefont {A.~A.}\ \bibnamefont
  {Mostofi}}, \bibinfo {author} {\bibfnamefont {J.~R.}\ \bibnamefont {Yates}},
  \bibinfo {author} {\bibfnamefont {I.}~\bibnamefont {Souza}}, \ and\ \bibinfo
  {author} {\bibfnamefont {D.}~\bibnamefont {Vanderbilt}},\ }\href {\doibase
  10.1103/RevModPhys.84.1419} {\bibfield  {journal} {\bibinfo  {journal} {Rev.
  Mod. Phys.}\ }\textbf {\bibinfo {volume} {84}},\ \bibinfo {pages} {1419}
  (\bibinfo {year} {2012})}\BibitemShut {NoStop}%
\bibitem [{\citenamefont {Zhou}\ \emph {et~al.}(2020)\citenamefont {Zhou},
  \citenamefont {Wan},\ and\ \citenamefont {Xu}}]{2020Zhou}%
  \BibitemOpen
\bibfield  {journal} {  }\bibfield  {author} {\bibinfo {author} {\bibfnamefont
  {Z.}~\bibnamefont {Zhou}}, \bibinfo {author} {\bibfnamefont {L.-L.}\
  \bibnamefont {Wan}}, \ and\ \bibinfo {author} {\bibfnamefont {Z.-F.}\
  \bibnamefont {Xu}},\ }\href {\doibase 10.1088/1751-8121/abb92b} {\bibfield
  {journal} {\bibinfo  {journal} {Journal of Physics A: Mathematical and
  Theoretical}\ }\textbf {\bibinfo {volume} {53}},\ \bibinfo {pages} {425203}
  (\bibinfo {year} {2020})}\BibitemShut {NoStop}%
\bibitem [{\citenamefont {Shastry}\ \emph {et~al.}(1993)\citenamefont
  {Shastry}, \citenamefont {Shraiman},\ and\ \citenamefont
  {Singh}}]{1993Shastry}%
  \BibitemOpen
  \bibfield  {author} {\bibinfo {author} {\bibfnamefont {B.~S.}\ \bibnamefont
  {Shastry}}, \bibinfo {author} {\bibfnamefont {B.~I.}\ \bibnamefont
  {Shraiman}}, \ and\ \bibinfo {author} {\bibfnamefont {R.~R.~P.}\ \bibnamefont
  {Singh}},\ }\href {\doibase 10.1103/PhysRevLett.70.2004} {\bibfield
  {journal} {\bibinfo  {journal} {Phys. Rev. Lett.}\ }\textbf {\bibinfo
  {volume} {70}},\ \bibinfo {pages} {2004} (\bibinfo {year}
  {1993})}\BibitemShut {NoStop}%
\bibitem [{\citenamefont {Shindou}\ \emph {et~al.}(2013)\citenamefont
  {Shindou}, \citenamefont {Matsumoto}, \citenamefont {Murakami},\ and\
  \citenamefont {Ohe}}]{2013Shindou}%
  \BibitemOpen
  \bibfield  {author} {\bibinfo {author} {\bibfnamefont {R.}~\bibnamefont
  {Shindou}}, \bibinfo {author} {\bibfnamefont {R.}~\bibnamefont {Matsumoto}},
  \bibinfo {author} {\bibfnamefont {S.}~\bibnamefont {Murakami}}, \ and\
  \bibinfo {author} {\bibfnamefont {J.-i.}\ \bibnamefont {Ohe}},\ }\href
  {\doibase 10.1103/PhysRevB.87.174427} {\bibfield  {journal} {\bibinfo
  {journal} {Phys. Rev. B}\ }\textbf {\bibinfo {volume} {87}},\ \bibinfo
  {pages} {174427} (\bibinfo {year} {2013})}\BibitemShut {NoStop}%
\bibitem [{\citenamefont {Engelhardt}\ and\ \citenamefont
  {Brandes}(2015)}]{Engelhardt2015}%
  \BibitemOpen
  \bibfield  {author} {\bibinfo {author} {\bibfnamefont {G.}~\bibnamefont
  {Engelhardt}}\ and\ \bibinfo {author} {\bibfnamefont {T.}~\bibnamefont
  {Brandes}},\ }\href {\doibase 10.1103/PhysRevA.91.053621} {\bibfield
  {journal} {\bibinfo  {journal} {Phys. Rev. A}\ }\textbf {\bibinfo {volume}
  {91}},\ \bibinfo {pages} {053621} (\bibinfo {year} {2015})}\BibitemShut
  {NoStop}%
\bibitem [{\citenamefont {Furukawa}\ and\ \citenamefont
  {Ueda}(2015)}]{2015Furukawa}%
  \BibitemOpen
  \bibfield  {author} {\bibinfo {author} {\bibfnamefont {S.}~\bibnamefont
  {Furukawa}}\ and\ \bibinfo {author} {\bibfnamefont {M.}~\bibnamefont
  {Ueda}},\ }\href {\doibase 10.1088/1367-2630/17/11/115014} {\bibfield
  {journal} {\bibinfo  {journal} {New Journal of Physics}\ }\textbf {\bibinfo
  {volume} {17}},\ \bibinfo {pages} {115014} (\bibinfo {year}
  {2015})}\BibitemShut {NoStop}%
\bibitem [{Note1()}]{Note1}%
  \BibitemOpen
  \bibinfo {note} {We note that in the expression for $\Omega _0({\protect \bm
  {K}})$ the summation over the band index excludes $\bar 0$ since
  $n=0,{\protect \bm {k}}={\protect \bm {K}}$ and $n=\bar 0,{\protect \bm
  {k}}={\protect \bm {K}}$ refer to the same ground state.}\BibitemShut {Stop}%
\bibitem [{\citenamefont {Kuhr}(2016)}]{2016Kuhr}%
  \BibitemOpen
  \bibfield  {author} {\bibinfo {author} {\bibfnamefont {S.}~\bibnamefont
  {Kuhr}},\ }\href {\doibase 10.1093/nsr/nww023} {\bibfield  {journal}
  {\bibinfo  {journal} {National Science Review}\ }\textbf {\bibinfo {volume}
  {3}},\ \bibinfo {pages} {170} (\bibinfo {year} {2016})}\BibitemShut {NoStop}%
\bibitem [{\citenamefont {Wu}\ \emph {et~al.}(2015)\citenamefont {Wu},
  \citenamefont {Taylor},\ and\ \citenamefont {Zaremba}}]{2015Wu}%
  \BibitemOpen
  \bibfield  {author} {\bibinfo {author} {\bibfnamefont {Z.}~\bibnamefont
  {Wu}}, \bibinfo {author} {\bibfnamefont {E.}~\bibnamefont {Taylor}}, \ and\
  \bibinfo {author} {\bibfnamefont {E.}~\bibnamefont {Zaremba}},\ }\href
  {\doibase 10.1209/0295-5075/110/26002} {\bibfield  {journal} {\bibinfo
  {journal} {{EPL} (Europhysics Letters)}\ }\textbf {\bibinfo {volume} {110}},\
  \bibinfo {pages} {26002} (\bibinfo {year} {2015})}\BibitemShut {NoStop}%
\bibitem [{\citenamefont {Anderson}\ \emph {et~al.}(2019)\citenamefont
  {Anderson}, \citenamefont {Wang}, \citenamefont {Xu}, \citenamefont {Venu},
  \citenamefont {Trotzky}, \citenamefont {Chevy},\ and\ \citenamefont
  {Thywissen}}]{2019Anderson}%
  \BibitemOpen
  \bibfield  {author} {\bibinfo {author} {\bibfnamefont {R.}~\bibnamefont
  {Anderson}}, \bibinfo {author} {\bibfnamefont {F.}~\bibnamefont {Wang}},
  \bibinfo {author} {\bibfnamefont {P.}~\bibnamefont {Xu}}, \bibinfo {author}
  {\bibfnamefont {V.}~\bibnamefont {Venu}}, \bibinfo {author} {\bibfnamefont
  {S.}~\bibnamefont {Trotzky}}, \bibinfo {author} {\bibfnamefont
  {F.}~\bibnamefont {Chevy}}, \ and\ \bibinfo {author} {\bibfnamefont {J.~H.}\
  \bibnamefont {Thywissen}},\ }\href {\doibase 10.1103/PhysRevLett.122.153602}
  {\bibfield  {journal} {\bibinfo  {journal} {Phys. Rev. Lett.}\ }\textbf
  {\bibinfo {volume} {122}},\ \bibinfo {pages} {153602} (\bibinfo {year}
  {2019})}\BibitemShut {NoStop}%
\bibitem [{\citenamefont {Tran}\ \emph {et~al.}(2017)\citenamefont {Tran},
  \citenamefont {Dauphin}, \citenamefont {Grushin}, \citenamefont {Zoller},\
  and\ \citenamefont {Goldman}}]{2017Tran}%
  \BibitemOpen
  \bibfield  {author} {\bibinfo {author} {\bibfnamefont {D.~T.}\ \bibnamefont
  {Tran}}, \bibinfo {author} {\bibfnamefont {A.}~\bibnamefont {Dauphin}},
  \bibinfo {author} {\bibfnamefont {A.~G.}\ \bibnamefont {Grushin}}, \bibinfo
  {author} {\bibfnamefont {P.}~\bibnamefont {Zoller}}, \ and\ \bibinfo {author}
  {\bibfnamefont {N.}~\bibnamefont {Goldman}},\ }\href {\doibase
  10.1126/sciadv.1701207} {\bibfield  {journal} {\bibinfo  {journal} {Science
  Advances}\ }\textbf {\bibinfo {volume} {3}},\ \bibinfo {pages} {1701207}
  (\bibinfo {year} {2017})}\BibitemShut {NoStop}%
\bibitem [{\citenamefont {Asteria}\ \emph {et~al.}(2019)\citenamefont
  {Asteria}, \citenamefont {Tran}, \citenamefont {Ozawa}, \citenamefont
  {Tarnowski}, \citenamefont {Rem}, \citenamefont {Fl{\"a}schner},
  \citenamefont {Sengstock}, \citenamefont {Goldman},\ and\ \citenamefont
  {Weitenberg}}]{2019Asteria}%
  \BibitemOpen
  \bibfield  {author} {\bibinfo {author} {\bibfnamefont {L.}~\bibnamefont
  {Asteria}}, \bibinfo {author} {\bibfnamefont {D.~T.}\ \bibnamefont {Tran}},
  \bibinfo {author} {\bibfnamefont {T.}~\bibnamefont {Ozawa}}, \bibinfo
  {author} {\bibfnamefont {M.}~\bibnamefont {Tarnowski}}, \bibinfo {author}
  {\bibfnamefont {B.~S.}\ \bibnamefont {Rem}}, \bibinfo {author} {\bibfnamefont
  {N.}~\bibnamefont {Fl{\"a}schner}}, \bibinfo {author} {\bibfnamefont
  {K.}~\bibnamefont {Sengstock}}, \bibinfo {author} {\bibfnamefont
  {N.}~\bibnamefont {Goldman}}, \ and\ \bibinfo {author} {\bibfnamefont
  {C.}~\bibnamefont {Weitenberg}},\ }\href {\doibase 10.1038/s41567-019-0417-8}
  {\bibfield  {journal} {\bibinfo  {journal} {Nature Physics}\ }\textbf
  {\bibinfo {volume} {15}},\ \bibinfo {pages} {449} (\bibinfo {year}
  {2019})}\BibitemShut {NoStop}%
\bibitem [{\citenamefont {Midtgaard}\ \emph {et~al.}(2020)\citenamefont
  {Midtgaard}, \citenamefont {Wu}, \citenamefont {Goldman},\ and\ \citenamefont
  {Bruun}}]{2020Midtgaard}%
  \BibitemOpen
  \bibfield  {author} {\bibinfo {author} {\bibfnamefont {J.~M.}\ \bibnamefont
  {Midtgaard}}, \bibinfo {author} {\bibfnamefont {Z.}~\bibnamefont {Wu}},
  \bibinfo {author} {\bibfnamefont {N.}~\bibnamefont {Goldman}}, \ and\
  \bibinfo {author} {\bibfnamefont {G.~M.}\ \bibnamefont {Bruun}},\ }\href
  {\doibase 10.1103/PhysRevResearch.2.033385} {\bibfield  {journal} {\bibinfo
  {journal} {Phys. Rev. Research}\ }\textbf {\bibinfo {volume} {2}},\ \bibinfo
  {pages} {033385} (\bibinfo {year} {2020})}\BibitemShut {NoStop}%
\bibitem [{\citenamefont {Petrov}\ \emph {et~al.}(2000)\citenamefont {Petrov},
  \citenamefont {Holzmann},\ and\ \citenamefont {Shlyapnikov}}]{Petrov2000}%
  \BibitemOpen
  \bibfield  {author} {\bibinfo {author} {\bibfnamefont {D.~S.}\ \bibnamefont
  {Petrov}}, \bibinfo {author} {\bibfnamefont {M.}~\bibnamefont {Holzmann}}, \
  and\ \bibinfo {author} {\bibfnamefont {G.~V.}\ \bibnamefont {Shlyapnikov}},\
  }\href {\doibase 10.1103/PhysRevLett.84.2551} {\bibfield  {journal} {\bibinfo
   {journal} {Phys. Rev. Lett.}\ }\textbf {\bibinfo {volume} {84}},\ \bibinfo
  {pages} {2551} (\bibinfo {year} {2000})}\BibitemShut {NoStop}%
\end{thebibliography}
\end{document}